\documentclass[onecolumn,english]{svjour3}
\usepackage[T1]{fontenc}
\usepackage[latin9]{inputenc}
\usepackage{color}
\usepackage{prettyref}
\usepackage{amsmath}
\usepackage{amssymb}
\usepackage{graphicx}

\makeatletter
\smartqed  

\makeatother

\usepackage{babel}
\begin{document}

\title{The Lorentz-Dirac equation in complex space-time}

\author{Mark Davidson}

\institute{Mark Davidson\at Spectel Research Corp., 807 Rorke Way, Palo Alto,
CA, USA 94303\\
Tel.: +001-650-424-0645\\
\email{mdavid@spectelresearch.com}\\
}

\date{Received: date / Accepted: date}
\maketitle
\begin{abstract}
A hypothetical equation of motion is proposed for Kerr-Newman particles.
It's obtained by analytic continuation of the Lorentz-Dirac equation
into complex space-time. A new class of ``runaway'' solutions are
found which are similar to zitterbewegung. Electromagnetic fields
generated by these motions are studied, and it's found that the retarded
(and advanced) times are multi-sheeted functions of the field points.
This leads to non-uniqueness for the fields. With fixed weighting
factors for these multiple roots, the solutions radiate. However,
position dependent weighting factors can suppress radiation and allow
non-radiating solutions. Motion with external forces are also considered,
and radiation suppression is possible there too. These results are
relevant for the idea that Kerr-Newman solutions provide insight into
elementary particles and into emergent quantum mechanics. They illustrate
a type of nascent wave-particle duality and complementarity in a purely
classical field theory. Metric curvature due to gravitation is ignored.

\end{abstract}

\section{Introduction\label{sec:Introduction}}

The Kerr-Newman (K-N) solutions of general relativity have been proposed
as classical models for elementary particles \cite{burinskii_kerr_2007,burinskii_dirac_2008,carter_global_1968,lynden-bell_magic_2003,newman_classical_2002,pekeris_electromagnetic_1987}.
Their gyromagnetic g factor is exactly 2, the same as the Dirac equation
\cite{carter_global_1968}. It's difficult to construct classical
models with this property. One would like to know the equation of
motion for such particles, but a rigorous derivation is lacking in
the relativity literature. This paper proposes an equation of motion
for them which is motivated by physical considerations, and then presents
some properties and solutions. The method exploits the complex manifold
techniques that have been prevalent in the literature on the K-N metric
\cite{adamo_null_2009,newman_complex_1973,newman_heaven_1976}. There
is recent renewed interest in the idea that quantum mechanics might
be emergent from a simple, possibly deterministic dynamic at the Planck
scale \cite{adler_quantum_2004,carroll_gravity_2004,markopoulou_quantum_2004,t_hooft_equivalence_1988,t_hooft_how_2001,t_hooft_determinism_2000,t_hooft_determinism_2005,t_hooft_entangled_2009,weinberg_collapse_2011}.
A dynamical theory for the K-N particle could yield clues for this
program. It has also been proposed that quantum mechanics might have
emerged when the universe was in the ideal fluid mechanical phase,
which is now believed to have occurred prior to about 3 microseconds
after the big-bang \cite{davidson_quark-gluon_2008,davidson_quark-gluon_2010},
and where K-N particles might possibly have been formed due to turbulent
density fluctuations in this fluid. 

The Lorentz-Dirac equation (LDE) applies to exactly point-like classical
charged particles in real Minkowski space. We study a complexified
version of it, and find new solutions which are similar to zitterbewegung
\cite{schroedinger_uber_1930,huang_zitterbewegung_1952,hestenes_zitterbewegung_1990}.
These have unusual properties. The electromagnetic fields are not
unique. Rather, an infinite number of possible fields can arise from
a single particle solution. The origin of this is the multi-sheeted
nature of the retarded time function. In fact, it has an infinite
number of Riemann sheets. The null condition can be satisfied by an
infinite number of different proper times for a given field point.
This is radically different than ordinary classical electrodynamics,
where only one unique single point can satisfy the retarded time condition.
The multi-sheeted structure is analyzed here using mainly analytical
but also numerical methods. The most general field solution is a linear
superposition over all the possible retarded times, with weighting
factors that are a countably infinite set of complex numbers (or even
position dependent functions), and which are not determined by the
theory. The situation bears a vague similarity to quantum mechanics,
and even the many-worlds interpretation, where a state vector is associated
with a particle and is required in order to specify the state of the
particle. Here the state vector corresponds to the set of complex
weights for the multitude of source points. The particle no longer
acts as a single point or string so far as the electromagnetic field
is concerned. This behavior is both intriguing but also very puzzling. 

The treatment here is done in complex Minkowski space, and the gravitational
interaction is ignored. A more rigorous and complete treatment would
require the use of the Debney-Kerr-Schild formalism \cite{debney_solutions_1969},
which puts strong restrictions on the dynamics of the solutions by
the extra requirement of the alignment of the Faraday tensor with
the directions of the null congruence. It is well known that a relativistic
quantum particle's position cannot be localized with only positive
energy states \cite{newton_localized_1949,wightman_localizability_1962},
and consequently a localized Schrödinger operator does not strictly
speaking exist. Consequently second quantized field theory has been
the central topic of modern particle physics for many years. However,
modified and improved approaches to single particle relativistic wave
equations have been proposed to remedy this situation \cite{prugovecki_consistent_1978,prugovecki_stochastic_1984,schroeck_quantum_1996,ali_conserved_1981},
and these ``stochastic phase space'' models may serve as a mathematical
bridge between the current theory and the Dirac equation for example.

\section{A brief overview of the Kerr-Newman solution and complex Minkowski
space}

The K-N particle is modeled as a point charge which is slightly displaced
from the real space-time hypersurface embedded in complex Minkowski
space (CM$^{\text{4}}$) with corrections to the metric tensor due
to gravity \cite{newman_metric_1965,newman_maxwells_1973,newman_complex_1973,newman_heaven_1976,burinskii_kerr_2007,burinskii_dirac_2008,lynden-bell_magic_2003,pekeris_electromagnetic_1987}.
The gravitational and electromagnetic fields are singular on a ring
and this endows the solution with string-like properties \cite{burinskii_gravitational_2011}.
When the angular momentum is taken to zero, the ring collapses to
a point singularity on the real Minkowski space as described by the
Reissner\textendash{}Nordström metric. When considered as a hypothetical
model for an elementary particle, the gravitational effects are minute
compared to electromagnetic ones \cite{burinskii_dirac_2008}, and
shall be ignored here. However, the incorporation of the full metric
calculation is extremely important for the ultimate importance of
the ideas presented here, especially considering the complications
and potential difficulties point out in \cite{burinskii_nonstationary_1995}
for accelerated charges. It is hoped that by focusing first on the
limiting electrodynamic theory the path to a full metric calculation
will be illuminated. The complex theory makes contact with reality
by ultimately considering the electromagnetic fields on the real space-time
hyperspace. The Riemann-Silberstein complex vector field is used (we
use boldface for 3-vectors and italics for 4-vectors and tensors)

\begin{equation}
\mathbf{W}=\mathbf{E}+i\mathbf{B}\label{eq:Riemann-Silberstein}
\end{equation}
or, in an equivalent covariant form, a complex Faraday tensor is introduced
\cite{newman_maxwells_1973}

\begin{equation}
W^{\mu\nu}=F^{\mu\nu}+i*F^{\mu\nu},\:*W=-iW\label{eq:Complex Faraday}
\end{equation}
where $F$ is the Faraday tensor and $*F$ is the Hodge dual. $W$
is anti self-dual (ASD). The energy and momentum densities are

\begin{equation}
\mathcal{E}_{el}=\frac{1}{2}\overline{{\normalcolor \mathbf{W}}}\cdot\mathbf{W},\:\mathbf{P}_{el}=\frac{i}{2}\mathbf{W}\times\overline{\mathbf{W}}\label{eq: energy and momentum for RS vector}
\end{equation}
The stress energy tensor on the real hyperspace is a function of the
real-valued physical Faraday tensor 

\begin{equation}
4\pi T^{\mu\nu}=F_{phys\:\lambda}^{\mu}F_{phys}^{\lambda\nu}+\frac{1}{4}g^{\mu\nu}F_{phys\alpha\beta}F_{phys}^{\alpha\beta}\label{eq: Stress energy tensor}
\end{equation}
\textcolor{black}{and this can be expressed in terms of the complex
Faraday tensor with the substitution }

\begin{equation}
F_{phys}^{\mu\nu}=Re\: W^{\mu\nu}\label{eq: Physical Faraday tensor}
\end{equation}
The absence of magnetic charge requires that $F_{phys}$ satisfies
the electromagnetic Bianchi identity, which is not automatically satisfied
if $F$ is complex valued in \prettyref{eq:Complex Faraday}, a problem
taken up in section 8. The static K-N particle is modeled as a point
charge located at a point in complex 3-space at $\mathbf{z_{0}}=\mathbf{x_{0}}+i\mathbf{b}$,
where $\mathbf{x_{0}}$ and $\mathbf{b}$ are real 3-vectors. However,
in many ways this object acts more like a string than a point charge
\cite{burinskii_gravitational_2011}. One introduces a complex \textcolor{black}{Coloumb}
potential 

\begin{equation}
\Phi=q/\sqrt{\left(\mathbf{z}-\mathbf{z}_{0}\right)^{2}}\label{eq: Complex_KN_potential}
\end{equation}
The complex vector inside the square root is simply squared and not
absolute-value-squared in this equation, so that the resulting $\psi$
is complex even for real values of $\mathbf{z}$. $\Phi$ is holomorphic
in each spatial coordinate and double-sheeted. It follows that $\nabla^{2}\Phi=0$
on the real hyperspace so long as $\mathbf{b}\neq0$. The prescription
for making physical sense out of this complex displacement is to take
the Riemann-Silberstein vector as given by the analytic continuation
of the electric field to complex values. Amazingly, this recipe yields
the correct electromagnetic field of the K-N solution with $\mathbf{W}=-\mathbf{\nabla}\Phi$.
The gradient is unambiguous because of the holomorphic property of
$\Phi$, and its real and imaginary parts yield the electric and magnetic
fields respectively. It also has two Riemann sheets because of the
square root. As there is no charge on the real hyperspace when $\mathbf{z}_{0}$
is complex, this leads to the so-called ``sourceless'' K-N solution.
The covariant version of this formula starts with a calculation of
the analytic continuation of the real Faraday tensor to its value
at the complex source point $\mathbf{z}_{0}$, $F_{c}^{\mu\nu}=\partial^{\mu}\Phi\delta^{0\nu}-\partial^{\nu}\Phi\delta^{0\mu}$,
where $\delta$ denotes the Kronecker delta. This complex tensor will
not be ASD automatically. In order to calculate the tensor $W$ we
must use \prettyref{eq:Complex Faraday} to project out the ASD part.
Then, on the real hypersapce, the real part will give the physical
Faraday tensor \prettyref{eq: Physical Faraday tensor}.

The metric tensor is found by solving the equations of general relativity
with a stress energy tensor given by \prettyref{eq: Stress energy tensor},
and by assuming that the metric can be expressed in Kerr-Schild coordinates
in the form \cite{debney_solutions_1969} $g_{\mu\nu}=\eta_{\mu\nu}-2Hk_{\mu}k_{\nu}$,
where $k$ is a null vector field, $H$ a scalar field, and $\eta$
is the Minkowski metric. The Kerr theorem provides a formula for the
most general geodesic and shear-free congruences on real or complexified
Minkowski space.

Introducing a discontinuity on the Riemann cut (which can be taken
as the disk bounded by the ring singularity) leads to a surface charge
and current density on this disk which can then be taken as a source
for the fields which are made single valued but discontinuous by this
procedure \cite{lynden-bell_electromagnetic_2004}. When considering
values for charge, mass, and angular momentum (q, m, and j) of elementary
particles, the gravitational metric is well-approximated by a Minkowski
metric except very near to the ring singularity \cite{burinskii_dirac_2008}.
The K-N solutions include the cases where the charge $q$ is complex.
This describes particles with both an electric and magnetic charge
\cite{calvani_latitudinal_1982,kasuya_exact_1982}. They also include
the case where the mass is negative. The total electromagnetic energy
stored in the static electric and magnetic fields of a K-N solution
is infinite \cite{lynden-bell_magic_2003}. Despite the fact that
the electromagnetic energy is infinite, the mass of the K-N particle
is finite, and is arbitrary and unrelated to the electromagnetic energy.
Consequently, we cannot interpret these solutions as purely electromagnetic
particles in the limit $G\rightarrow0$. Something must account for
the cancellation of the infinite electrostatic energy to yield a finite
mass, so the Poincaré stress problem for the classical electron persists
for the static K-N particle as well. Along these lines, an interesting
mechanism to render the energy finite by introducing Higgs fields
and modifying the purely electromagnetic K-N stress-energy tensor
was proposed in \cite{burinskii_regularized_2010}. The K-N solutions
have event horizons whenever the mass is sufficiently large such that
(in Planck units) $m^{2}>(j/m)^{2}+q^{2}$ \cite{burinskii_dirac_2008,carter_global_1968}.
For the case of elementary particles, event horizons would not be
present, and consequently the ring singularity would be naked. This
violates the cosmic censorship hypothesis \cite{penrose_question_1999},
but whether this is sufficient reason to abandon consideration of
such K-N particles is unclear at the present time. In the regularized
solution of \cite{burinskii_regularized_2010} there is no singularity.

\section{The Lorentz-Dirac equation (LDE)}

The LDE for a single particle is \cite{dirac_classical_1938,jackson_classical_1999,plass_classical_1961,rohrlich_classical_2007,teitelboim_classical_1980} 

\begin{equation}
m_{0}a^{\mu}=F_{ext}^{\mu}+m_{0}\tau_{0}\left(\dot{a}^{\mu}+a^{\lambda}a_{\lambda}v^{\mu}\right)\label{eq:Lorentz-Dirac equation}
\end{equation}
where $F_{ext}^{\mu}$ is any external force, $\tau_{0}=\frac{2q_{0}^{2}}{3m_{0}c^{3}}$,
and c=1, $a$ and $v$ are the proper acceleration and velocity respectively,
and $\tau$ is the proper time. Einstein summation convention is assumed,
and dot-product notation will be used for 4-vector dot products as
well, so that $a^{\lambda}a_{\lambda}=a\cdot a$. The metric signature
is (+,-,-,-). We use the notation $A^{[\mu}B^{\nu]}=A^{\mu}B^{\nu}-A^{\nu}B^{\mu}$.
Boldface type will be used to denote 3-vectors, and their dot products
will be as usual. For time-like particles, the equation is supplemented
by subsidiary constraints on the proper acceleration and velocity
which imply that in the rest frame, the acceleration 4-vector is space-like

\begin{equation}
v\cdot v=1,\; v\cdot a=0\label{eq: Velocity Acceleration constraints}
\end{equation}
The vast literature on this equation is reasonably in agreement on
two points. First, if we consider exactly point particles and not
extended quasi-particles, then the general consensus is that the LDE
is correct in the sense that straightforward electromagnetic analysis
leads to it. In fact it can be derived on purely geometrical grounds
\cite{ringermacher_intrinsic_1979}. Secondly, this equation has one
of two possible unphysical properties - either it has runaway solutions,
or it has pre-acceleration. There is a tendency to replace the LDE
in practical calculations by an approximate equation developed first
by Landau and Lifshitz \cite{jackson_comment_2007,landau_classical_1951}
which is suitable for non point-like quasi-particles and free of pathologies.
Since the K-N solutions are obtained from a monopole in complex space-time,
it is reasonable to consider the LDE and not the quasi-particle equation
for describing them. The term $m\tau_{0}\dot{a}$ in \prettyref{eq:Lorentz-Dirac equation}
is called the Schott term. Its origin can be understood from the following
equation for the total momentum of the charged particle including
electromagnetic momentum \cite{teitelboim_splitting_1970}. $p^{\mu}=mv^{\mu}-\frac{2}{3}e^{2}a^{\mu}$
. The term $m\tau_{0}a^{2}v$ in \prettyref{eq:Lorentz-Dirac equation}
is the force caused by emitted radiation. In the method developed
in this paper, the runaway solutions lead to interesting behavior,
and therefore we study them in detail. It may turn out that this behavior
will also be considered unphysical after further analysis, but in
the meantime they are novel and seem worthy of study.

\section{Runaway solutions }

The most famous runaway solution of the LDE is \cite{rohrlich_classical_2007}
for motion in the z-t plane

\begin{equation}
v^{\mu}=(cosh(Ae^{\tau/\tau_{0}}+B),\:0,\:0,\: sinh(Ae^{\tau/\tau_{0}}+B))^{\mu}
\end{equation}
where $A$ and $B$ are constants. Let's call these type I runaway
solutions. Here the focus will be on another type of runaway solution.
Consider the LDE, and Look for a solution which has vanishing acceleration
squared in the absence of force

\begin{equation}
a^{\lambda}a_{\lambda}=0\label{eq:radiation free}
\end{equation}
then \prettyref{eq:Lorentz-Dirac equation} becomes very simply $a^{\mu}=\tau_{0}\dot{a}^{\mu}$
with solution

\begin{equation}
a^{\mu}(\tau)=e^{\tau/\tau_{0}}a^{\mu}(0),\: a^{\lambda}(0)a_{\lambda}(0)=0\label{eq: Acceleration for complex type II}
\end{equation}

\begin{equation}
v^{\mu}(\tau)=\tau_{0}e^{\tau/\tau_{0}}a^{\mu}(0)+u^{\mu}\label{eq: Velocity for complex type II}
\end{equation}

\begin{equation}
z^{\mu}(\tau)=\left(\tau_{0}\right)^{2}e^{\tau/\tau_{0}}a^{\mu}(0)+u^{\mu}\tau+\Omega^{\mu}\label{eq:coordinate solution}
\end{equation}
where $\Omega^{u}$ and $u^{\mu}$ are constant 4-vectors. For time-like
particles, there are no non-trivial solutions which satisfy the constraints
\prettyref{eq: Velocity Acceleration constraints}. In complex space-time,
there are non-trivial solutions of this type, and these shall be the
focus of this paper. These solutions shall be referred to as type
II. Note that the Laplace transform in $\tau$ of a type II solution
has a simple pole whereas the type I solutions have an infinite number
of poles. So they are considerably simpler. Runaway solutions, dismissed
as unphysical for over 100 years, have posed a profound challenge
for classical electromagnetism. Note that both type I and II solutions
are entire functions of $\tau$ if it is considered as a complex variable.
They are also analytic functions of the constants of the motion $a^{\mu}(0)$,
$\Omega^{\mu}$, and $u^{\mu}$.

\section{Analytic continuation of free particle equation and solutions}

The Free-particle LDE and its solutions can be analytically continued
into the complex domain. We will let $\tau$, $z^{\mu}$ $a^{\mu}(0)$,
$\Omega^{\mu}$, and $u^{\mu}$ all become complex variables. Considering
first only the dependence on $\tau$, the position of the particle
becomes not a curve but rather a two dimensional surface embedded
in complex Minkowski space or CM$^{\text{4}}$. This is because $\tau$
has both a real and imaginary part. Some of the literature has concluded
that the particle acts like a string \cite{burinskii_gravitational_2011}.
There is much merit in this string point of view, considering that
the fields produced by the K-N solution are singular on a circular
ring. Suppose we try and enforce that we have a point particle equation
of motion by requiring that the particle move along a curve in complex
space-time by constraining it to move along a curve in the complex
$\tau$ plane. Let the curve be parametrized by some real variable
$s$ so that the curve of the particle in complex Minkowski space
is given by $z^{\mu}(\tau(s))$. The analytic continuation of the
LDE and its solutions to complex $\tau$ are uniquely determined by
their real hyperspace forms. We could interpret this as meaning that
any analytic function $\tau(s)$ yields a possible solution for the
charged particle's motion due to self-forces while moving in complex
space-time. One trajectory stands out in importance, and this is the
one for which $z^{0}$, the time component, stays real along the trajectory.
We think of this as the effective trajectory of the particle, while
acknowledging the underlying string-like nature of the solution too
\cite{burinskii_gravitational_2011}. Despite this non-uniqueness,
one particular analytic continuation will be studied in detail here
because of its similarity to zitterbewegung. Starting with all real
values for the various initial parameters, set $\tau(s)=s$ for s
real and ranging from $-\infty$ to $+\infty$ and then we analytically
continue this curve to the curve $\tau(s)=is$ along the imaginary
$\tau$ axis. An analytic transformation which accomplishes this is
the following continuous function as $\lambda$ vary from 0 to 1.

\begin{equation}
\tau(\lambda,s)=s+(i-1)\lambda s,\:\lambda\in\left[0,1\right]\label{eq:analytic continuation}
\end{equation}
Actually, the reader may prefer to avoid trajectories altogether at
this stage by simply making a change of complex variables $\tau=is$,
but for the remainder of this section we shall restrict consideration
to the trajectory mapped out by letting $s$ and $z^{0}$ vary along
the real axis. Additional analytic continuations of the initial values
$\Omega^{\mu},\: u^{\mu},\: a(0)^{\mu}$ will also be made in order
to obtain a physically sensible solution. Applying \prettyref{eq:analytic continuation}
results in $\tau=is$ and consequently the LDE becomes 

\begin{equation}
-m_{0}\frac{d^{2}z}{ds^{2}}^{\mu}=F_{ext}^{\mu}+im_{0}\tau_{0}\left(\frac{d^{3}z}{ds^{3}}^{\mu}-\left(\frac{d^{2}z}{ds^{2}}\right)^{2}\frac{dz}{ds}^{\mu}\right)\label{eq: Pure analytic continuation}
\end{equation}
We wish to interpret $s$ as a new scaled proper time for the motion
of the particle in the real hyperspace. This will allow us to fix
some free parameters of the theory. Notice that then the inertial
mass has the wrong sign though. A physical particle must have a positive
inertial mass. The K-N metric has both positive and negative mass
solutions, so we therefore choose $m_{0}$ to be negative. Note that
if $m_{0}<0$ then $\tau_{0}$ changes sign too, so we have to take
care with the signs. At the risk of confusion we will nevertheless
continue to let $\tau_{0}$ denote the positive value. We have then,
in the absence of external force

\begin{equation}
m\frac{d^{2}z}{ds^{2}}^{\mu}=im\tau_{0}\left(\frac{d^{3}z}{ds^{3}}^{\mu}-\left(\frac{d^{2}z}{ds^{2}}\right)^{2}\frac{dz}{ds}^{\mu}\right),\quad m=-m_{0}\:>0\label{eq: Zero force LD equation in s}
\end{equation}
The inertial and gravitational mass must be equal. We assume for
the time being that this is possible with these assumptions, but to
prove it would require solving for the gravitational metric which
is difficult for the cases of interest considering \cite{burinskii_nonstationary_1995}.
The change in sign for the mass might be due to some form of self
interaction. Considering the type II runaway solutions \prettyref{eq:coordinate solution}
we have 

\begin{equation}
a_{s}^{\mu}(s)=d^{2}z^{\mu}/ds^{2}=e^{-is/\tau_{0}}a_{s}^{\mu}(0),\quad a_{s}^{\lambda}(0)a_{s\lambda}(0)=0\label{eq: a(s)}
\end{equation}

\begin{equation}
v_{s}^{\mu}(s)=dz^{\mu}/ds=i\tau_{0}e^{-is/\tau_{0}}a_{s}^{\mu}(0)+V^{\mu}\label{eq: v(s)}
\end{equation}

\begin{equation}
z_{s}^{\mu}(s)=-\tau_{0}^{\,2}e^{-is/\tau_{0}}a_{s}^{\mu}(0)+V^{\mu}s+\Omega^{\mu}\label{eq: z(s)}
\end{equation}
and $V$ and $\Omega$ are constant 4-vectors, with $V$ real and
time-like.

\subsection{Circular rotating solution for free particles}

Consider analytic continuation in the constant 4-vector $a_{s}(0)$
to the following pair of possible null vectors

\begin{equation}
a_{\pm}(0)=\delta_{a}(0,1,\pm i,0)\label{eq: a(0)  for x-y orbit}
\end{equation}
which will be seen to have opposite chirality. Although $\delta_{a}$
can be a complex constant, by choice of the zero point for $s$ we
can absorb its phase, and so we take it to be real and positive. These
null vectors lead to a localized oscillation in the complex space
rather than a runaway solution. For future reference, the following
formulas are easily verified

\begin{equation}
a_{\pm}(0)\cdot x=-\delta_{a}r_{\bot}e^{\pm i\varphi}\label{eq: azimuthal phase result}
\end{equation}

\begin{equation}
\left(Re\: a_{\pm}(0)\right)\cdot\left(Im\: a_{\pm}(0)\right)=0
\end{equation}
where $r_{\bot}$and $\varphi$ are the radial and azimuthal coordinates
for the 3-vector part of a real 4-vector $x$ in cylindrical coordinates.
The most general solution of this type can be found by noting that
an arbitrary complex 3-vector can be written $\mathbf{a_{s}(0)}=\mathbf{A}+i\mathbf{B}$
where $\mathbf{A}$ and $\mathbf{B}$ are real. Then, $\mathbf{a_{s}(0)^{2}}=0\Rightarrow\mathbf{A}\cdot\mathbf{B}=0$
and $\mathbf{A^{\mathrm{2}}}=\mathbf{B}^{2}$. These are all obtained
from \prettyref{eq: a(0)  for x-y orbit} the by a 3-rotation of $a_{\pm}(0)$.
We let $\Omega^{\mu}$ have a small imaginary part to give it the
properties of a spinning particle as dictated by the K-N static solution
$\Omega^{\mu}=x_{0}^{\,\mu}+ib^{\mu}$, where $b^{\mu}$ is real and
spacelike, and in the ``average rest frame'' where $V^{\mu}=(\gamma,0,0,0)$,
$b^{\mu}=(0,\overrightarrow{b})$, and $\gamma$ is a real constant.
In this example we have chosen to take $\vec{b}$ along the 3-direction
to maintain symmetry. It might be oriented in other directions too,
but these shall not be considered here. The magnitude of $\vec{b}$
is the angular momentum divided by mass for the K-N particle. For
an electron, it would be $\hslash/2m_{e}$. We shall allow values
of $\gamma\neq1$ as this allows us to interpret the real part of
$z_{s}$ as a true particle trajectory with proper time given by s.
Note that $v_{s}\cdot v_{s}=\gamma^{2}\neq1$, so this is a departure
from ordinary classical mechanics. However, note also that with this
complex acceleration, there is no rest frame for the particle which
can be reached from the Laboratory frame with a real-valued Lorentz
transformation. Writing out the real and imaginary parts in full (with
$\delta_{a}$ real and positive and for $s$ real) 

\begin{equation}
Re(z_{\pm}^{\mu})=-\tau_{0}^{2}\delta_{a}(0,\: cos(s/\tau_{0}),\:\pm sin(s/\tau_{0}),\:0)^{\mu}+V^{\mu}s+x_{0}^{\,\mu}
\end{equation}

\begin{equation}
Im(z_{\pm}^{\mu})=-\tau_{0}^{2}\delta_{a}(0,\:-sin(s/\tau_{0}),\:\pm cos(s/\tau_{0}),\:0)^{\mu}+b^{\mu}
\end{equation}
The real coordinates are the closest we can come describing this object
as a normal particle. It is misleading to think that the particle
is really a simple charge moving in a circle, just as the semiclassical
interpretation of zitterbewegung in the Dirac equation as a circular
particle motion is also a gross over-simplification. The proper velocity
of the real trajectory is (for real values of $s$)

\begin{equation}
\frac{d}{ds}Re(z_{\pm}^{\mu})=-\tau_{0}\delta_{a}(0,\:-sin(s/\tau_{0}),\:\pm cos(s/\tau_{0}),\:0)^{\mu}+V^{\mu}
\end{equation}
and this must satisfy

\begin{equation}
\frac{d}{ds}Re(z_{\pm}^{\mu})\frac{d}{ds}Re(z_{\pm\mu})=\gamma^{2}-(\tau_{0}\delta_{a})^{2}=1\label{eq: equation for gamma and delta}
\end{equation}
which is required if this is to be interpreted as a time-like particle
with $s$ its proper time. Therefore we must have $|\delta_{a}|=\sqrt{\gamma^{2}-1}/\tau_{0}$.
It also follows for these solutions that

\begin{equation}
\left(\frac{d}{ds}z_{\pm}^{\mu}\right)a_{\pm\mu}=0\:,\mathrm{and}\:\left(\frac{d}{ds}Re(z_{\pm}^{\mu})\right)Re(a_{\pm\mu})=0
\end{equation}

\begin{equation}
\left(dz/ds\right)^{2}=V^{\mu}V_{\mu}=\gamma^{2}
\end{equation}
It is clear that the two solutions circulate with opposite chirality
in the real hyperspace. The determination of a value to assign to
$\gamma$ is not obvious. In order to fix it, we identify the oscillation
with zitterbewegung as described by the Dirac equation \cite{dirac_principles_1978,schroedinger_uber_1930,huang_zitterbewegung_1952,barut_zitterbewegung_1981,hestenes_zitterbewegung_1990}
which satisfies $\omega_{z}=2m_{obs}/\hslash$ where $m_{obs}$ is
the positive physical mass of the observed particle, and the one to
be used in the Dirac equation. Laboratory time is given by

\begin{equation}
t=Re\: z^{0}=V^{0}s=\gamma s
\end{equation}
Expressing the oscillation in terms of this time, and equating it
with the zitterbewegung angular frequency gives

\begin{equation}
e^{-is/\tau_{0}}=e^{-it/\gamma\tau_{0}}=e^{-i\omega_{z}t}=e^{-i(2m_{obs}/\hbar)t}
\end{equation}

\begin{equation}
\frac{1}{\gamma\tau_{0}}=\omega_{z}=\frac{2m_{obs}}{\hbar},\:\tau_{0}=\frac{2q^{2}}{3m}
\end{equation}
Considering \prettyref{eq: equation for gamma and delta}, special
relativity requires that the mechanical energy of the rotating point
mass be $m_{obs}=\gamma m=\gamma|m_{0}|$ so that the observed mass
must be different from the original static K-N mass in magnitude and
sign. On combining these equations one finds that $\gamma$ is independent
of mass and given by $\gamma^{2}=\frac{3\hslash}{4q^{2}}$. Introducing
the fine structure constant $\alpha=\frac{e^{2}}{\hslash}$ we have
for a particle with charge $e$

\begin{equation}
\gamma=\sqrt{\frac{3}{4\alpha}}=10.1378991576698\:\pm1.66\times10^{-8}\label{eq: Zitterbewegung frequency condition for gamma}
\end{equation}
The uncertainty reflects the experimental uncertainty in $\alpha$.
We've obtained this precise value of $\gamma$ by requiring that $s$
be the proper time of the projected real motion, and although this
is a natural assumption, it's not absolutely mandatory and one could
consider other positive values for $\gamma$. We can interpret the
solution as a particle with an internal clock which has twice the
frequency as mandated by quantum mechanics - the de Broglie clock
\cite{de_broglie_researches_1924} - which is the same as the result
for the Dirac equation. In the average rest frame $Re\: z^{\mu}$
describes a circular orbit of radius $\tau_{0}^{2}\delta_{a}$ and
angular frequency $\omega_{z}$. The laboratory speed, determined
by the real part of $z$ is $\tau_{0}^{2}\mid\delta_{a}\mid\omega$.
If we accept the usual arguments regarding zitterbewegung, then this
would be the speed of light. This is only approximately true in our
case. Solving \prettyref{eq: equation for gamma and delta} for $\delta$
yields $\delta_{a}=\sqrt{\gamma^{2}-1}/\tau_{0}$. The radius of oscillation
is then approximately half of the (reduced) Compton wavelength $\lambda_{C}=2\gamma\tau_{0}$

\begin{equation}
r=\tau_{0}^{2}\delta_{a}=\tau_{0}\sqrt{\gamma^{2}-1}=\frac{\lambda_{C}\sqrt{\gamma^{2}-1}}{2\gamma}
\end{equation}
so that the speed as viewed in the mean rest frame of the particle
model for an electron is independent of mass and given by

\begin{equation}
r\omega_{z}=\sqrt{\gamma^{2}-1}/\gamma=(0.995123206731709\:\pm1.60\times10^{-11})c
\end{equation}
 In most models for zitterbewegung the speed is taken to be exactly
c and the radius is given by \cite{schroedinger_uber_1930,barut_zitterbewegung_1981,hestenes_zitterbewegung_1990}
$r_{S}=\frac{\hslash}{2m_{obs}}=\gamma\tau_{0}$. This value, originally
due to Schrödinger, agrees with our radius to better than 1\%. The
angular momentum due to this circular motion alone may be calculated
to be $L=\frac{\hbar}{2}\frac{\gamma^{2}-1}{\gamma^{2}}$. This is
slightly smaller than the spin of the electron by about 1\%. In addition
to this purely mechanical angular momentum, there is also a contribution
from the electromagnetic field angular momentum. It is hoped that
this extra contribution will make up the deficit and yield a total
angular moment of $\hbar/2$. These results are in qualitative agreement
with zitterbewegung for the Dirac equation, but the speeds are subluminal
here, and the real hyperspace projection of the particle's world line
describes a time-like particle. Underlying this apparent real motion
though is a periodic null motion in complex space-time which was obtained
by an analytic continuation of the LDE. The self-accelerating runaway
solutions have been transformed by this procedure into localized oscillatory
solutions, much like the zitterbewegung of the Dirac equation \cite{dirac_principles_1978,huang_zitterbewegung_1952,sidharth_revisiting_2008,hestenes_zitterbewegung_1990}.
The physical existence of zitterbewegung has been experimentally verified
through numerous spectroscopic calculations which need to include
the Darwin term \cite{bjorken_relativistic_1998} to agree with experimental
measurements of the hyperfine splitting of atomic lines in spectroscopy,
and also more directly by observing resonant behavior in electron
channeling \cite{guoanere_experimental_2008}. We've had to introduce
mass renormalization, change the sign of the mass, and allow complex
valued acceleration to have a plausible theory. The reward for these
radical assumptions is that we have arrived at a solution to a purely
classical equation which describes phenomenon that we actually observe
in nature and which is normally associated with relativistic quantum
mechanics. Moreover, the runaway solutions have been eliminated, without
imposing the usual boundary condition that the acceleration vanish
in the infinite future. In our case the acceleration endures forever,
as does the zitterbewegung for a free Dirac particle. This explanation
for zitterbewegung has a problem if neutrinos are included in the
discussion, since if the mass of the neutrino is exactly zero, then
$\tau_{0}$ would be undefined, but if its mass is non-zero, as is
now believed, then $\tau_{0}=0$ and there would be no zitterbewegung.
Many other solutions can be obtained from these ones by applying the
full symmetry group of transformations for the complex LDE. This group
includes as a subgroup the complex ten parameter Poincare group \cite{newman_complex_1973,streater_pct_2000}
together with parity symmetry as well as complex conjugation of the
initial acceleration 4-vector. Time reversal is not a symmetry of
the LDE, presumably because a preferred direction in time is introduced
in its derivation by allowing only the retarded fields to self-interact
with the particle. Wheeler-Feynman electrodynamics \cite{wheeler_interaction_1945-1}
in contrast is time symmetric. There is earlier literature showing
a connection between the LDE and zitterbewegung as in \cite{browne_electron_1970-1},
but the relationship to the current theory is not clear at this time
to the author.

\subsection{space-time loop solution}

Another class of solutions can be obtained by letting $a_{s}^{\mu}(0)=\delta_{03}(1,0,0,1)+\delta_{12}(0,1,\pm i,0)$
which is a null vector. This describes an additional oscillatory behavior
in the z-t variables. If $a_{s}\cdot V=0$ then we must have $V^{\mu}=0$
if the mean 4-velocity is real and time-like, and we then get a closed
loop in complex space-time. The motion is governed by arbitrary complex
constants $\delta_{03}$ and $\delta_{12}$. 

\begin{equation}
z_{s}(s)=-\tau_{0}^{\,2}e^{-is}\left[\delta_{03}(1,0,0,1)+\delta_{12}(0,1,\pm i,0)\right]+\Omega
\end{equation}
Taking the real part of this yields circular oscillation in the 1-2
plane synchronized with linear oscillation in the 0-3 plane. As the
time oscillates in this solution, it violates causality, although
on a very tiny scale. The static K-N solutions are known to have time-like
closed loop geodesics \cite{morris_wormholes_1988,morris_wormholes_1988-1}.
This loop solution is qualitatively similar to a virtual particle
loop in a Feynman diagram, and although we will not be considering
these solutions further, they are pointed out here because of their
possible relevance to emergent quantum mechanics. The most general
complex null 4-vector $a_{s}(0)$ can be written as $a_{s}(0)=(\pm\sqrt{\mathbf{k\cdot k}},\:\mathbf{k})$,
for arbitrary complex 3-vector $\mathbf{k}$.

\section{Li\'{e}nard-Wiechert potentials and the fields}

The Li\'{e}nard-Wiechert potentials provide a procedure for calculating
the electromagnetic fields produced by a particle. In ordinary electromagnetism
they are \cite{jackson_classical_1999}

\begin{equation}
A^{\mu}(x)=\frac{qv^{\mu}(\tau)}{v(\tau)\cdot(x-z(\tau))}\mid_{\tau=\tau_{r}}\label{eq: Lienard-Wiechert potential}
\end{equation}
For example, consider the static K-N case, for which the source point
is $z(\tau)=(\tau,0,0,ib)$. The solutions for the retarded root is
$\tau_{r}=x^{0}-\sqrt{\left(\mathbf{x}-i\mathbf{b}\right)^{2}}$.
Substitution into \prettyref{eq: Lienard-Wiechert potential} then
yields \prettyref{eq: Complex_KN_potential}. We obtain the ring singularity
from what looks mathematically at least like a single source point
in the complex space-time. It is for this reason that we draw a distinction
between the source points and the singularity points. Here the source
seems to be a monopole located at $z(\tau_{r})$, but the field is
singular when the field point $\mathbf{x}$ is on a ring in real Minkowski
space. Note that if we reparamatrize the curve by $z(\tau)=z(\tau(s))$
then we have

\begin{equation}
\frac{dz^{\mu}(\tau)/d\tau}{dz(\tau)/d\tau\cdot(x-z(\tau))}=\frac{dz^{\mu}(\tau(s))/ds}{dz(\tau(s))/ds\cdot(x-z(\tau(s)))}\label{eq: Lienard Wiechert sum}
\end{equation}
where $\tau_{r}$ is the proper time which satisfies the null root
equation $\left(x-z(\tau)\right)\cdot\left(x-z(\tau)\right)=0$ or
$\left(x-z(\tau(s))\right)\cdot\left(x-z(\tau(s))\right)=0$. After
analytic continuation the coordinates are expressed in terms of the
new variable s and the complex valued potential $A$ is used to calculate
a complex Faraday tensor $F$ and its dual $*F$ from which the anti
self-dual tensor $W$ can be calculated from \prettyref{eq:Complex Faraday}.
If the particle trajectory is time-like in the (real valued) case
of standard electromagnetic theory, then there are exactly two solutions
to the null root equation, one in the past (retarded time) and one
in the future (advanced time). When we analytically continue these
equations into the complex plane, the null root equation becomes for
complex $z$ and $s$ and real $x$

\begin{equation}
\left(x-z(s)\right)^{\mu}\left(x-z(s)\right)_{\mu}=0\label{eq:Null root equation}
\end{equation}
This condition in the complex case is fundamentally different. Both
the real and imaginary parts of \prettyref{eq:Null root equation}
must vanish. We are considering here the situation where $z$ is parametrized
along a curve by a real parameter $s$, and the null root equation
will quite likely have no solutions at all for real $s$. In general,
the null root equation will have solutions which are not on the path
of the particle. Nevertheless, these solutions are obtained by analytic
continuation from ordinary solutions, and so they are valid ones.
This is an important - indeed profound - difference as compared to
the usual case. The set of solutions $\left\{ s_{i}\right\} $ to
\prettyref{eq:Null root equation} can be quite large, even infinite
as we shall see below. Some subset will be causal, but even this subset
can be infinite. Why shouldn't we include a superposition of several
different roots at the same time? The solutions for the electromagnetic
field are not unique and may be characterized by the number of roots
included in the solution, by the particular values of these roots,
and by their respective weights in the sum. For single-root solutions
we have, letting $v^{\mu}=dz^{\mu}/ds$, 

\begin{equation}
A^{\mu}(x)=\frac{qv_{s}^{\mu}(s)}{v_{s}(s)\cdot(x-z(s))},\: s\in\left\{ s_{i}\right\} \label{eq: Single root soluton}
\end{equation}
If there are more than one retarded solution, then there is no reason
to exclude multiple root solutions. For an N root solution we have 

\begin{equation}
A^{\mu}(x)=\sum_{j=1}^{N}w_{j}\frac{qv_{s}^{\mu}(s_{j})}{v_{s}(s_{j})\cdot(x-z(s_{j}))},\: s_{j}\in\left\{ s_{i}\right\} ,\;\sum_{j=1}^{N}w_{j}=1\label{eq: superposition of source points}
\end{equation}
These 4-potentials are in general complex. The rule for calculating
the complex Faraday tensor \prettyref{eq:Complex Faraday} from them
is $W^{\mu\nu}=\partial^{[\mu}A^{\nu]}+i*\partial^{[\mu}A^{\nu]}$
and the physical Faraday tensor on the real hyperspace $x$ is the
real part of this $F_{physical}=Re\: W$. There is no assurance that
the electromagnetic Bianchi identity will be satisfied, and so magnetic
currents can appear from this prescription, but they can be suppressed
by the methods of section 8. The weight factors $w_{j}$ are arbitrary
except that we would expect that they sum to one, and we may want
to include only causal retarded times in the sum. For the time being
we shall assume that these weighting factors are constants, but later
on we shall consider the possibility that they are dependent on position.
These multi-root solutions would look to an observer capable of measuring
the electromagnetic fields produced by them like a particle made up
of multiple constituents, but they are really analytic continuations
associated with a single particle trajectory. So we see we have a
whole family of possible solutions. Already in the literature there
is evidence for this non-uniqueness. It is exploited in Wheeler-Feynman
electrodynamics \cite{wheeler_interaction_1945-1} by allowing a sum
of advanced and retarded potentials to give solutions to Maxwell's
equation, although this is a fairly trivial case in comparison. The
author has wondered whether these multi-root solutions might be related
to the quark model for hadrons. They could look like a particle made
up of multiple constituent point particles if probed electromagnetically.
But these ``particles'' might be hard to separate because they are
related to a single particle trajectory in the complex space, with
their positions linked together by the requirement that they must
lie on different Riemann sheets of a single root equation, perhaps
providing a mechanism for quark confinement. The Li\'{e}nard-Wiechert
equations can be used for analytic continuation without specifying
a mathematical form for the charge current in the complex space-time.
We can think of this analytic continuation as proceeding in the following
way to understand why summing over the different roots is reasonable.
Suppose we start with real-space Li\'{e}nard-Wiechert potential for
a single root, and write it in the following way:

\begin{equation}
A^{\mu}(x)=\sum_{j}w_{j}\frac{qv^{\mu}(\tau_{r})}{v(\tau_{r})\cdot(x-z(\tau_{r}))}
\end{equation}
(We haven't done anything yet, since $\sum w_{j}=1$). But now we
analytically continue each term in the sum to a different Riemann
sheet, and thus obtain a different value for the retarded time for
each term in the sum. This is a generalization of the usual analytic
continuation for analytic functions in mathematics. The result is
a field which is produced by complex source point charges at a countable
set of different complex valued retarded times. The maximum number
of source points is the number of Riemann sheets of the null time
function.

\section{Calculation of retarded proper times for the type II runaway solutions}

\subsection{Null time solutions for the circular rotating solutions}

The Null root equation that must be satisfied for solutions to \prettyref{eq:Null root equation}
for trajectories described by \prettyref{eq: z(s)} is, for a real
field point $x^{\mu}$ 

\begin{equation}
\left(x+\tau_{0}^{\,2}e^{-is/\tau_{0}}a_{s}(0)-Vs-\Omega\right)^{2}=0\label{eq:Null Root for Z solutions}
\end{equation}
The solution set $\left\{ s_{j}\right\} $ will be implicit functions
of $x$, $a_{s}(0)$, $u$, and $\Omega$, and in general will not
be real and consequently $z(s_{j})$ will not lie on the particle's
real-time trajectory in the complex space-time. This is a consequence
of relying on analytic continuation to guide us into the complex space,
and it reflects the string-like morphology of the K-N solutions \cite{burinskii_gravitational_2011}.
Expanding the square we obtain 

\begin{equation}
\left(x-Vs-\Omega\right)^{2}+2\tau_{0}^{\,2}e^{-is/\tau_{0}}a_{s}(0)\cdot\left(x-Vs-\Omega\right)=0
\end{equation}
We have $V=(\gamma,0,0,0)$, $\Omega\cdot V=\Omega\cdot a_{s}(0)=V\cdot a_{s}(0)=0$,
$z^{0}(s)=\gamma s$, $a_{s}(0)$ is given by \prettyref{eq: a(0)  for x-y orbit},
and $\Omega=(0,0,0,ib)$. This corresponds to a K-N particle with
spin oriented in the 3 direction and experiencing complex rotation
in the 1-2 plane. The center of the orbit is at rest. The null condition
becomes 

\begin{equation}
\left(x-\Omega\right)^{2}-2x^{0}\gamma s+\left(\gamma s\right)^{2}=-2\tau_{0}^{\,2}e^{-is/\tau_{0}}a_{s}(0)\cdot x\label{eq: null root equation}
\end{equation}
By factorizing we may rewrite this equation in the form

\begin{equation}
\left(s-C_{A}\right)\left(s-C_{R}\right)=De^{-is/\tau_{0}}\label{eq: simple root equation}
\end{equation}
where $C_{A}$, $C_{R}$, and $D$ are known functions of x. 

\begin{equation}
C_{A}=\left(+2x^{0}\gamma+\sqrt{\left(2x^{0}\gamma\right)^{2}-4\gamma^{2}R^{2}}\right)/2\gamma^{2}
\end{equation}

\begin{equation}
C_{R}=\left(+2x^{0}\gamma-\sqrt{\left(2x^{0}\gamma\right)^{2}-4\gamma^{2}R^{2}}\right)/2\gamma^{2}
\end{equation}

\begin{equation}
D_{\pm}=-2\tau_{0}^{\,2}a_{\pm}(0)\cdot x/\gamma^{2}=2\tau_{0}^{\,2}(\delta_{a}r_{\bot}e^{\pm i\varphi})/\gamma^{2}\label{eq:D azimuthal phase}
\end{equation}

\begin{equation}
R^{2}=(x-\Omega)^{2},\;\Omega=(0,0,0,ib)
\end{equation}

Let us consider the right-handed case to be specific

\begin{equation}
a_{s}(0)=a_{+}(0)=\delta_{a}(0,1,i,0)
\end{equation}
It's not possible to solve \prettyref{eq: simple root equation} in
closed form, unless one uses a proposed generalization of the Lambert
function \cite{scott_general_2006}, but this generalization has not
been sufficiently analyzed in the literature to be useful yet. We
can proceed numerically by a method of successive approximation. Start
with $C_{A}$ and $C_{R}$ set to zero, so that the equation then
becomes:

\begin{equation}
s^{2}=De^{-is/\tau_{0}}\label{eq: first approximation to root}
\end{equation}
This can be solved in terms of Lambert functions \cite{corless_lambert_1996}. 

\begin{equation}
s=-2i\tau_{0}W_{n}(\pm i\sqrt{D}/(2\tau_{0})),\: n\in\mathbb{Z}
\end{equation}
Where $W_{n}$ is the nth branch of the Lambert function. There are
an infinite number of Riemann sheets, and an infinite number of roots.
Starting with this solution, we can analytically continue \prettyref{eq: simple root equation}
to the correct values for $C_{A}$ and $C_{R}$ using numerical means.
Thus we can conclude that there are likely to be an infinite number
of solutions, unless the analytic continuation results in infinitely
many of these different starting approximations coalescing to the
same value. Numerical simulation of this analytic continuation suggests
that this does not happen, and they indicate the existence of an infinite
number of solutions. These solutions correspond to different Riemann
sheets of the solution $s$ when viewed as a function of $x$. The
situation is illustrated in figure 1.

\begin{figure}
\includegraphics{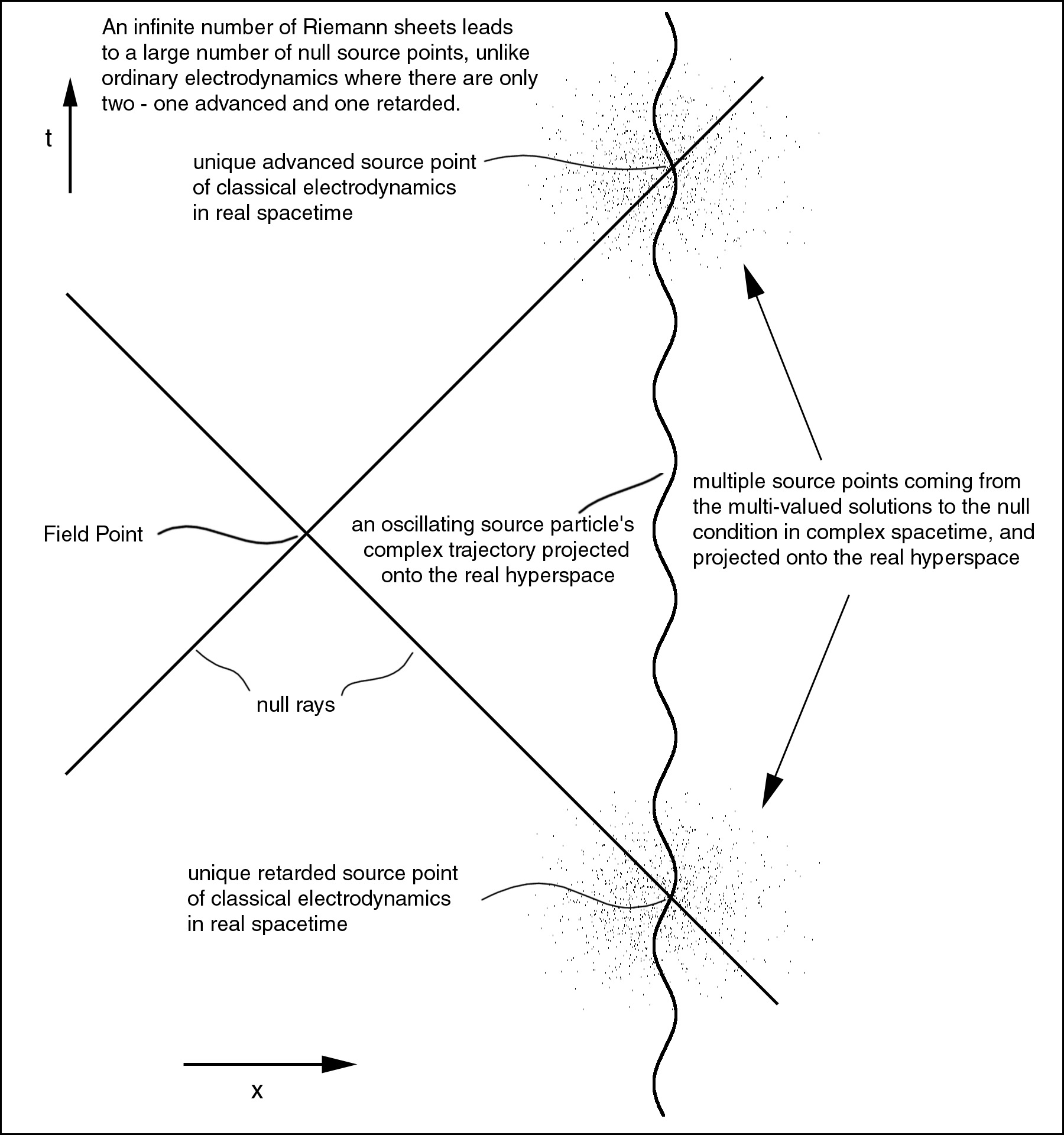}

\caption{\label{fig:one-column-figure-1}Qualitative illustration of multiple
source points for Li\'{e}nard-Wiechert potentials}
\end{figure}

\subsection{Calculation of the roots in the asymptotic field limit}

It is important to examine if these solutions radiate energy, which
would render them unphysical. Even if they do radiate, however, it
might be possible to find radiation backgrounds (such as the zero
point radiation \cite{boyer_random_1975,de_la_pena_quantum_1996})
with which these solutions could be in equilibrium, so that they would
absorb as much energy on the average as they radiated, and leave the
statistical distribution of radiation unchanged. In order to study
the radiation from the source, we make the standard approximation
for large $r=|\mathbf{x}|$, and let $t=x^{0}$. We assume that $\left|\mathbf{z}\right|\ll\left|\mathbf{x}\right|$
and therefore we make the standard far-field approximation with $x^{\mu}$
real

\begin{equation}
\left(x-z(s)\right)^{2}=0\thickapprox\left(t-z^{0}\right)^{2}-(\mathbf{x^{2}}\mathrm{-2}\mathbf{x}\cdot\mathbf{z})
\end{equation}

\begin{equation}
t-z^{0}=t-\gamma s\thickapprox\pm(r-\hat{\mathbf{r}}\cdot\mathbf{z)}=\pm\left(r+\tau_{0}^{2}(\mathbf{a}_{s}(0)\cdot\mathbf{x})e^{-is/\tau_{0}}/r-\hat{\mathbf{r}}\cdot\mathbf{\Omega}\right)
\end{equation}
For the retarded solution $s\uparrow$we choose the + sign and obtain

\begin{equation}
\gamma s\uparrow-(t-r+\hat{\mathbf{r}}\cdot\mathbf{\Omega})+\tau_{0}^{2}(\mathbf{a_{s}}(0)\cdot\mathbf{\hat{r}})e^{-is\uparrow/\tau_{0}}=0
\end{equation}

\begin{equation}
A\equiv(t-r+\hat{\mathbf{r}}\cdot\mathbf{\Omega}),\: B\equiv\tau_{0}^{2}(\mathbf{a_{s}}(0)\cdot\mathbf{\hat{r}}),\:\gamma s\uparrow-A+Be^{-is\uparrow/\tau_{0}}=0
\end{equation}
This equation can be solved for $s\uparrow$ again in terms of Lambert
functions 

\begin{equation}
s\uparrow=A/\gamma-i\tau_{0}W_{n}\left(-\frac{iBe^{-iA/\gamma\tau_{0}}}{\gamma\tau_{0}}\right),\: n\in\mathbb{Z}
\end{equation}

\begin{equation}
s\uparrow_{n}=(t-r+\hat{\mathbf{r}}\cdot\mathbf{\Omega})/\gamma-i\tau_{0}W_{n}\left(-\frac{i\tau_{0}^{2}(\mathbf{a_{s}}(0)\cdot\mathbf{\hat{r}})e^{-i(t-r+\hat{\mathbf{r}}\cdot\mathbf{\Omega})/\gamma\tau_{0}}}{\gamma\tau_{0}}\right)\label{eq: retarded asymptotic root}
\end{equation}

The solution for the advanced root $s\downarrow$ is

\begin{equation}
s\downarrow_{n}=(t+r-\hat{\mathbf{r}}\cdot\mathbf{\Omega})/\gamma-i\tau_{0}W_{n}\left(+\frac{i\tau_{0}^{2}(\mathbf{a_{s}}(0)\cdot\mathbf{\hat{r}})e^{-i(t+r-\hat{\mathbf{r}}\cdot\mathbf{\Omega})/\gamma\tau_{0}}}{\gamma\tau_{0}}\right)\label{eq: Advanced asymptotic root}
\end{equation}

The large argument behavior of the Lambert functions are given by
the series expansion \cite{corless_lambert_1996}

\begin{align}
W_{k}(z) & =log(z)+2\pi ik-log(log(z)+2\pi ik)\nonumber \\
 & +\sum_{j=0}^{\infty}\sum_{m=1}^{\infty}c_{jm}log^{m}\left(log(z)+2\pi ik\right)\left(log(z)+2\pi ik\right)^{-j-m}\label{eq: Lambert Taylor series for large z}
\end{align}

There are an infinite number of solutions as n is an arbitrary integer.
The Poynting vector may be calculated from \prettyref{eq: energy and momentum for RS vector}.
The potential is then given by \prettyref{eq: superposition of source points}.
The analytic structure of the multi-sheeted Lambert functions are
of critical importance in understanding these solutions. We follow
the branch-cut conventions of \cite{corless_lambert_1996}. $W_{0}(z)$
is special as it has a branch point at $-1/e$ with a cut $\{z:-\infty<z\leq-1/e\}$.
$W_{1}(z)$ and $W_{-1}(z)$ each have two cuts given by $\{z:-\infty<z\leq-1/e\}$
and $\{z:-\infty<z\leq0\}$. All other branches have a single cut
$\{z:-\infty<z\leq0\}$. The function $W_{k+1}$ is obtained from
the function $W_{k}$ by analytically continuing it through its branch
cut in the counterclockwise direction, mimicking the complex log function
\cite{corless_lambert_1996}. Imagine holding $t$ and $\hat{\mathbf{r}}$
fixed, and analytically continuing the solutions by increasing $r$.
Recall \prettyref{eq: azimuthal phase result}, so that there is an
exponential factor in the argument of $W_{n}$ of the form

\begin{equation}
phase\: factor=e^{\pm i\varphi-i(t-r+\hat{\mathbf{r}}\cdot\mathbf{o})/\gamma\tau_{0}}\label{eq: phase factor for branch}
\end{equation}
Every time its phase increases by $2\pi$, the Lambert function moves
to a different branch. The asymptotic field, which determines whether
or not the particle radiates, depends on how we take $r$ to $\infty.$
If we hold $\varphi$ fixed we get one result, if we hold $r/\gamma\tau_{0}\pm\varphi$
fixed we get another.

\subsubsection{Holding $\left((r-t)/\gamma\tau_{0}\pm\varphi\right)$ fixed as $r\rightarrow\infty$ }

If we hold the phase \prettyref{eq: phase factor for branch} fixed
as we take $r$ to $\infty$ then $n$ does not change with $r$ and
the Lambert function stays on the same branch as we approach the $r\rightarrow\infty$
limit. In this case we have the result

\begin{equation}
s\uparrow_{n}=(t-r+\hat{\mathbf{r}}\cdot\mathbf{\Omega})/\gamma+O(1)
\end{equation}

\begin{equation}
s\downarrow_{n}=(t+r-\hat{\mathbf{r}}\cdot\mathbf{\Omega})/\gamma+O(1)
\end{equation}
This is reasonable behavior for an oscillating charge, and these solutions
would lead to radiation.

\subsubsection{Holding $\varphi$ and $t$ fixed as $r\rightarrow\infty$ }

This gives a different result for the asymptotic behavior. We have
the following recursion relations for analytic continuations starting
at $r$ and ending at $r+2\pi\gamma\tau_{0}$with $t$ and $\varphi$
held fixed

\begin{equation}
s\uparrow_{n}(r+2\pi\gamma\tau_{0})=s\uparrow_{n+1}(r)-2\pi\tau_{0},\; n\notin\left\{ -1,0\right\} 
\end{equation}

\begin{equation}
s\downarrow_{n}(r+2\pi\gamma\tau_{0})=s\downarrow_{n-1}(r)+2\pi\tau_{0},\; n\notin\left\{ 0,1\right\} 
\end{equation}
When $n=-1,\,0,\,\mathrm{or}\,1$ the behavior depends on the magnitude
of the argument of $W_{n}$. For example,

\begin{equation}
s\uparrow_{0}(r+2\pi\gamma\tau_{0})=s\uparrow_{0}(r)-2\pi\tau_{0},\;\mathrm{if}\:\left|\frac{\tau_{0}^{2}(\mathbf{a}(0)\cdot\mathbf{\hat{r}})}{\gamma\tau_{0}}\right|<1/e
\end{equation}

\begin{equation}
s\uparrow_{-1}(r+2\pi\gamma\tau_{0})=s\uparrow_{1}(r)-2\pi\tau_{0},\;\mathrm{if}\:\left|\frac{\tau_{0}^{2}(\mathbf{a}(0)\cdot\mathbf{\hat{r}})}{\gamma\tau_{0}}\right|<1/e
\end{equation}
and so in this case if the argument of $W_{n}$ is below the critical
value of $1/e$ and the solution starts on the principal branch with
$n=0$, it will stay on the principal branch for all $r$. In this
case $z(s)$ is single-valued. Otherwise, as $r\rightarrow\infty,$the
branch function $n(r)$ will approach infinity in a linear staircase
fashion. In general $n(r)$ will be either fixed at $0$ or a step-like
function of $r$ which grows linearly (expect for $n=-1$) in $r$
. If r is held fixed and $t$ varied, a similar behavior ensues in
the variables $r-t$ for the retarded case and $r+t$ for the advanced
case. In order to calculate the radiation, the asymptotic fields for
large $r$ and fixed $t$ are required. The Lambert function satisfies
the equation $z=W_{n}(z)e^{W_{n}(z)}$, and this is useful for asymptotic
analysis. For large $n$ we have from \prettyref{eq: Lambert Taylor series for large z}

\begin{equation}
W_{n}(\Gamma)=\left(log(\Gamma)+2\pi in\right)\:-log(log(\Gamma)+2\pi in)\:+o(log(n))
\end{equation}

\begin{equation}
\Gamma\equiv\Lambda e^{-i(t-r+\hat{\mathbf{r}}\cdot\mathbf{o})/\gamma\tau_{0}},\;\Lambda\equiv-\frac{i\tau_{0}(\mathbf{a}(0)\cdot\mathbf{\hat{r}})}{\gamma}
\end{equation}

\begin{equation}
s\uparrow_{n}=(t-r+\hat{\mathbf{r}}\cdot\mathbf{o})/\gamma-i\tau_{0}\left[(log(\Gamma)+2\pi in)-log(log(\Gamma)+2\pi in)\right]+o(log(n)))
\end{equation}
but $n$ grows in a staircase fashion as $n\sim(r-t)/2\pi\gamma\tau_{0}+n_{0}$
if we analytically continue along a radial line, for some constant
$n_{0}$, so 

\begin{equation}
s\uparrow_{n}=i\tau_{0}log(r)+o(log(r))\label{eq: strange retarded behavior}
\end{equation}
This is perplexing behavior. Now consider the case where $n=0$ and
the argument of $W_{0}$ is less in magnitude than the critical value
of $1/e$. Let us perform a Taylor's series expansion keeping only
the first term $W_{0}(z)=z+o(z)$, for small $\left|z\right|$.

\begin{equation}
s\uparrow_{0}\approx(t-r+\hat{\mathbf{r}}\cdot\mathbf{o})/\gamma-\frac{\tau_{0}^{2}(\mathbf{a}(0)\cdot\mathbf{\hat{r}})e^{-i(t-r+\hat{\mathbf{r}}\cdot\mathbf{o})/\gamma\tau_{0}}}{\gamma}
\end{equation}
This solution will radiate energy away, but the solution stays on
the same Riemann sheet as we take the limit. So we have an infinite
number of Riemann sheets for the solutions. Another example of multiple
sheet structure in the literature occurs when there are multiple K-N
particles present\cite{burinskii_wonderful_2005}.

\section{Suppressing of radiation and magnetic currents with weighting factors
that are functions of x}

When the previous results for the retarded proper times are used to
calculate the asymptotic fields, the author believes based on extensive
analysis that the system will either radiate electromagnetic energy
or else have some other unphysical feature. Moreover, and more disturbingly,
the asymptotic field depends on the path of analytic continuation
to large $r$. Adding together even an infinite number of roots with
complex weighting constants does not seem to alter this conclusion.
In order to suppress the radiation, we therefore consider allowing
the weighting factors $w_{i}$ in \prettyref{eq: superposition of source points}
to be functions of the field point $x$. So long as these functions
are analytic (or at least holomorphic) in the coordinates, the resulting
fields will be too. It is convenient to rewrite \prettyref{eq: superposition of source points}
using \prettyref{eq: v(s)} as follows

\begin{equation}
A^{\mu}(x)=A_{Coul}^{\mu}(x)+A_{Rad}^{\mu}(x)
\end{equation}

\begin{equation}
A_{Coul}^{\mu}(x)=\sum_{j=1}^{N}w_{j}(x)\frac{q\gamma\delta_{0}^{\mu}}{v(s_{j}(x))\cdot(x-z(s_{j}(x)))},\:\sum_{j=1}^{N}w_{j}(x)=1
\end{equation}

\begin{equation}
A_{Rad}^{\mu}(x)=\sum_{j=1}^{N}w_{j}(x)\frac{qi\tau_{0}e^{-is_{j}(x)/\tau_{0}}a_{\pm}^{\mu}(0)}{v(s_{j}(x))\cdot(x-z(s_{j}(x)))}
\end{equation}
and where $s_{j}(x)$ are the proper-time null roots. The potential
$A_{Coul}$ determines the Coulomb field and the potential $A_{Rad}$
determines the radiation field. In order to calculate the radiation
field, we must perform the following steps following our procedure
$W^{\mu\nu}(x)=\partial^{[\mu}A^{\nu]}+i(*\partial^{[\mu}A^{\nu]})$,
where the real part of $W$ gives the physical Faraday tensor. Note
the important fact that

\begin{equation}
*\partial^{[\mu}A^{\nu]}=i\partial^{[\mu}A^{\nu]}\Rightarrow W^{\mu\nu}(x)=0
\end{equation}
In other words, if $\partial^{[\mu}A^{\nu]}$ is self-dual, then the
resulting $W$ field is zero. In order for there to be radiation from
a current source, the fields produced must fall to zero as $1/r$
for large $r$. Consequently, the only term that can contribute radiation
is $A_{Rad}$. Note that $A_{Rad}$ has the simple form of a scalar
function of $x$ times a constant 4-vector $a_{\pm}^{\mu}(0)$. We
have for large $r$, $\partial^{[\mu}A^{\nu]}=\partial^{[\mu}A_{Rad}^{\nu]}+o(\frac{1}{r})$.
If the $A_{Rad}$ happens to be self-dual, then there will be no radiation.
But the $w_{j}(x)$ are now assumed to be arbitrary functions, and
we can choose them such that this is the case. The following potential
functions are self dual in this sense 

\begin{equation}
A_{SD}(x)=f_{\pm}(x^{3}\mp x^{0},x^{1}\pm ix^{2})a_{\pm}(0)
\end{equation}
where $f_{\pm}$ are arbitrary holomorphic functions of their arguments.
So to cloak radiation for + or - chirality, we must choose the weighting
functions to satisfy

\begin{equation}
\sum_{j=1}^{N}w_{j}(x)\frac{qi\tau_{0}e^{-is_{j}(x)/\tau_{0}}}{v(s_{j}(x))\cdot(x-z(s_{j}(x)))}=f_{\pm}(x^{3}\mp x^{0},x^{1}\pm ix^{2});\quad\sum_{j=1}^{N}w_{j}(x)=1\label{eq: Non-radiating condition}
\end{equation}
Since two equations must be satisfied, at least two roots must be
included in the solution. Denote two such roots by index 1 and 2.
The equations can be solved simply with the result 

\begin{equation}
\left(\begin{array}{c}
w_{1}\\
w_{2}
\end{array}\right)=\frac{1}{C_{1}-C_{2}}\left(\begin{array}{c}
f_{\pm}-C_{2}\\
-f_{\pm}+C_{1}
\end{array}\right)
\end{equation}
where the two $C$ values in this formula are the functions multiplying
the respective weight functions $w$ in \prettyref{eq: Non-radiating condition}.
To this solution might be added by superposition any combination of
the weighting functions which satisfy the associated homogeneous equations.
Thus, there are a large number of non-radiating solutions. When the
cloaking conditions are satisfied, we can write

\begin{equation}
W^{\mu\nu}=\partial^{[\mu}A_{Coul}^{\nu]}+i(*\partial^{[\mu}A_{Coul}^{\nu]})
\end{equation}
and the radiation term vanishes. The weighting functions may be multi-sheeted
too. The resulting ``cloaked'' field will still generally be multi-sheeted,
except in the far-field where it will be just the single-valued Coulomb
field. The near field will contain electromagnetic currents which
depend on the weighting functions, and will not be sourceless in general. 

We can eliminate magnetic currents by again exploiting position dependent
weighting functions. In the compact notation of exterior derivatives
and differential forms (in 4 dimensions), the absence of magnetic
4-current on the real hyperspace requires that $d(ReW)=0$, the electromagnetic
Bianchi identity. This expression is linear in the weighting functions
$w_{i}$ and their first derivatives. In order to solve it, four additional
weighted root terms would be required in general, because the dual
of this equation is the magnetic current 1-form which has four independent
components. The solution is more complicated than the suppression
of radiation, since it involves derivatives of the weighting functions,
but it is linear and first order, and there should be solutions. In
general relativity, the most studied electromagnetic systems are sourceless
ones. We can also generate these by instead imposing the conditions
$dW=0$.

Surprisingly, with only three roots one can enforce that the fields
of the oscillating particle in real space-time are exactly the same
as the static K-N solution. This can be achieved by adding an additional
contraint equation for the vector potential 1-forms $A_{coul}=A_{Kerr-Newman}$
which is linear in the weighting functions, but does not involve any
derivatives, and so only a single extra root is required (because
the vector potentials have only one nonzero component). This is possible
because both terms have the same asymptotic behavior for large r.
This solution of course does not have any magnetic currents, and in
fact it is sourceless. This suggests that such solutions do exist.
Given this, a perturbative approach to study small deviations about
the K-N solution might be interesting. It also shows that stringlike
singularities can appear in this theory too.

The K\_N solution has infinite electrostatic energy, but there may
exist other finite energy solutions for the fields here, perhaps involving
still more roots, which may even be single valued, but have a more
complex near-field structure.

One question that naturally arises is what is the state of minimum
electrostatic energy? This will depend on the number of roots included
in the sum, assuming that there are some solutions for which the energy
is not infinite. Relevant to the question of emergence, the quantum
mechanical wave equations for free particles are closely related to
non-radiating electromagnetic sources \cite{davidson_quantum_2007}.

\section{Electromagnetic self energy and field singularities}

The electromagnetic field energy will be infinite if the complex vector
potential has a singularity. One way a singularity can occur is if
the denominator of any of the terms in the sum \prettyref{eq: Lienard Wiechert sum}
vanish which requires simultaneously

\begin{equation}
\left(x-z(s)\right)^{2}=0,\:\mathrm{and}\:\frac{\partial}{\partial s}\left(x-z(s)\right)^{2}=0
\end{equation}
and which for the present case becomes (together with \prettyref{eq: Single root soluton})

\begin{equation}
2s-(C_{A}+C_{R})=\frac{-i}{\tau_{0}}De^{-is/\tau_{0}}\label{eq: singularity equation}
\end{equation}
This equation can be solved for $s$ again with Lambert functions

\begin{equation}
s=\frac{1}{2}\left[\left(C_{A}+C_{R}\right)-2i\tau_{0}W_{n}\left(\frac{De^{-i\left(C_{A}+C_{R}\right)/2\tau_{0}}}{2\tau_{0}^{2}}\right)\right]\label{eq: Singular Lambert solution}
\end{equation}
From dividing \prettyref{eq: simple root equation} and \prettyref{eq: singularity equation}
we get a quadratic equation

\begin{equation}
2\left(s-\frac{\left(C_{A}+C_{R}\right)}{2}\right)+\frac{i}{\tau_{0}}\left(\left(s-\frac{\left(C_{A}+C_{R}\right)}{2}\right)^{2}-\left(\frac{C_{A}-C_{R}}{2}\right)^{2}\right)=0\label{eq: Singular quadratic}
\end{equation}
and these two equations must be simultaneously satisfied, and solving
this problem is very difficult. If the singularity occurs on the physical
Riemann sheet, the electromagnetic self energy will be infinite. However,
if all the singularities are off the physically realized Riemann sheets,
then this energy could be finite. It is an open problem to elucidate
the cases and calculate the self energies if any are finite. A particularly
simple case occurs when $(\mathbf{x}-\mathbf{\mathbf{\Omega}})^{2}=0$
for then $C_{A}=C_{R}=x^{0}/\gamma=C$. This is the condition for
points lying on the singularity ring of the static K-N particle, and
then the equations simplify to

\begin{equation}
s=C+2i\tau_{0}=C-2i\tau_{0}W_{n}\left(\pm\frac{i}{2\tau_{0}}\sqrt{\frac{D}{e^{iC/\tau_{0}}}}\right)
\end{equation}
so in order to have this singularity on this K-N ring, we must have
$W_{n}\left(\pm\frac{i}{2\tau_{0}}\sqrt{\frac{D}{e^{iC/\tau_{0}}}}\right)=-1$.
The only solutions to $W_{n}(y)=-1$ are $y=-1/e$, $n=0,1.$ There
is always some azimuthal angle $\varphi$ that gives the right phase
to the argument of $W_{n}$ by giving the phase of $D$ from \prettyref{eq:D azimuthal phase}
the correct value. Then magnitude of $D$ is determined uniquely if
this singularity is to occur. For the elementary particle values,
this magnitude value is not satisfied, and therefore there are no
field singularities on the K-N ring. There might be singularities
at other field points though, and perhaps there might even be singularities
along string-like curves in space. It seems possible that there could
be root solutions that don't have any field singularities on the physical
sheet too.

\section{Motion with weak external forces applied}

In an external electromagnetic field, we take the equation of motion
to be

\begin{equation}
m\frac{d^{2}z}{ds^{2}}^{\mu}=qF_{ext}^{\mu\nu}\frac{dz_{\upsilon}}{ds}+im\tau_{0}\left(\frac{d^{3}z}{ds^{3}}^{\mu}-\left(\frac{d^{2}z}{ds^{2}}\right)^{2}\frac{dz}{ds}^{\mu}\right)\label{eq: LDE with external F}
\end{equation}
The external Faraday tensor $F_{ext}$ here is real on the real hyperspace,
but is generally complex in the rest of the complex space-time by
analytic continuation. We assume that the fields are weak and slowly
varying on the time scale of $\tau_{0}.$ This reduces to simply the
Lorentz equation if we ignore the Schott and radiation terms, which
now have a factor of $i$ in front of them. This factor is rather
mysterious, and it seems questionable. The equation gives results
that we actually observe in nature, at least qualitatively, and it
is much better than the ordinary LDE equation in this regard, so we
will persist with this line of reasoning, even though the strange
form of the equation raises a number of questions and must be considered
preliminary. Consider the simplified case where we ignore the radiation
term. The equation then becomes

\begin{equation}
m\frac{d^{2}z}{ds^{2}}^{\mu}=qF_{ext\,\nu}^{\mu}\frac{dz^{\nu}}{ds}+im\tau_{0}\frac{d^{3}z}{ds^{3}}^{\mu}
\end{equation}
In this formula, the mass $m$ is the observed mass divided by $\gamma$.
It is natural to make a guiding center approximation to this. The
particle moves in a tight high-frequency oscillation perturbed by
the weak field. One finds quite generally that the Riemann structure
of the solution is a distortion of the Riemann structure of the simpler
pure zitterbewegung case described above. Thus, in the guiding center
approximation there will generally be an infinite number of retarded
time points, and the cloaking mechanism for shielding radiation can
be used for these too. Therefore, the radiation can be greatly suppressed
and in some cases even made zero.

\subsection{Motion in a 3D harmonic oscillator central force}

\begin{equation}
m\frac{d^{2}z}{ds^{2}}^{j}=-kz^{j}+im\tau_{0}\left(\frac{d^{3}z}{ds^{3}}^{j}-\left(\frac{d^{2}z}{ds^{2}}\right)^{2}\frac{dz}{ds}^{j}\right),\: j=1,2,3
\end{equation}

There will be exact solutions with $a\cdot a=0$ as follows. Then
the equation has solutions of the form $\mathbf{\mathrm{z\mathrm{^{j}(s)=}z^{j}(0)e^{i\omega s}}}$
with $\mathbf{z(0)\cdot z(0)}=0$. We can take $\mathrm{\mathbf{z}}\mathbf{(0)}$
to be proportional to either of the null vectors $\mathbf{a_{\pm}(0)}$
3-vectors for example for some arbitrary spin axis. Then the frequency
equation becomes $\tau_{0}\omega^{3}+\omega^{2}-k/m=0$. There are
three roots to this which are approximately $\omega=\pm\sqrt{k/m},-1/\tau_{0}$.
A class of exact solutions are given by a linear superposition of
these

\begin{equation}
\mathbf{z(s)}=\mathbf{a_{\pm}(0)}\sum_{j=1}^{3}c_{j}e^{i\omega_{j}s},\:\mathrm{and}\: z^{0}(s)=\gamma s
\end{equation}
for arbitrary complex constants $c_{j}$. These solutions are more
complicated than the single-frequency zitterbewegung solutions. Nevertheless,
we expect that they will certainly have multiple null roots for the
radiation calculation. Therefore, the same cloaking technique that
was proposed to eliminate radiation in the zitterbewegung case can
be applied in the present circumstance to reduce or eliminate the
radiation from these harmonic oscillator solutions as well.

\section{Particle morphology and complementarity}

We have seen that even if we take the interpretation that the motion
of the solution to the Lorentz-Dirac equation in complex space-time
is a one dimensional curve, as ordinary classical mechanics would
suggest, we still find that the source points in the Li\'{e}nard-Wiechert
potentials are not lying on this same curve, and in fact they are
distributed on a subset of the space-time world sheet of a string
as in Burinskii \cite{burinskii_gravitational_2011}. The static K-N
solution has a string-like ring singularity. In the present case,
the locus of singular points might also be string-like, but it is
difficult to show this because it requires solutions to the null root
equation for field points which are in the near-field of the particle.
For a singularity to occur, additional conditions must be satisfied
as in section 9 above. The set of field singularities may still lie
on a string, but this is not obvious to the author at least. One cannot
prove this without detailed numerical calculations or perhaps with
clever analytical methods. Moreover, as we are considering superpositions
of multiple roots, the solution might involve more than one string.
So even if we try and treat the motion as if it were one-dimensional
in the complex space-time, we are likely forced into a string-like
picture for the source points of the field, and possibly for the singularities
of the field as well. 

Complex space-time is an eight dimensional manifold. Imagine that
all the particles of the universe each have a string-like world sheet
in this manifold. Imagine too a kind of Mach's principle for this
space, namely, that given any clump of matter consisting of large
numbers of such particles, that the averaged coordinate values of
all this matter are narrowly clustered around a 4-dimensional real
hyperspace, and that this hyperspace has been singled out from all
others by this fact - a spontaneous symmetry breaking. This is our
world of Minkowski space. Then objects consisting of large numbers
of such particles can be described by real-valued positions and times.
This elevates the projection to real values of the complex coordinates
of individual particles to a distinguished meaning when being observed
by macroscopic experimental apparatus. 

The duality between pointlike and string-like behavior is similar
in spirit to Bohr's complementarity philosophy for quantum phenomena,
although we are considering a purely classical field theory here,
and it's also very suggestive from the point of view of emergent quantum
mechanics. Just as quantum mechanics presents us with paradoxical
complentarity, so too the present theory seems to do the same.

\section{Conclusion}

The picture that emerges from this model is of a particle with a peculiar
internal oscillation, and whose electromagnetic fields are dependent
on a set of complex weighting functions. The theory eliminates runaway
solutions from classical electromagnetism and replaces them with oscillating
solutions that look similar to zitterbewegung. These solutions oscillate
in time, which gives a tangible model for the de Broglie clock. They
allow a cloaking mechanism for radiationless accelerated motion, both
in the self-oscillating free particle case and the central harmonic
force case. Some might have string-like singularities, or even have
finite electromagnetic energy. The non-uniqueness is vaguely similar
to wave-particle duality. 

The calculation of the gravitational metric along the lines of \cite{debney_solutions_1969,burinskii_nonstationary_1995}
is an extremely important task needed to build confidence and add
value to the ideas presented here.
\begin{acknowledgements}
The author acknowledges Alexander Burinskii for informative correspondence
and the open source groups supporting the Octave and Maxima computer
languages along with the Wolfram-Alpha website which have been used
in the course of this work.
\end{acknowledgements}

\end{document}